\documentclass[%
 aip,
cp,  
 amsmath,amssymb,
 reprint,%
]{revtex4-2}

\usepackage{graphicx}
\usepackage{dcolumn}
\usepackage{bm}

\usepackage[utf8]{inputenc}
\usepackage[T1]{fontenc}
\usepackage{mathptmx} 

\begin{document}

\title{Explore Simpler Eigenmarking: \\
Quantum Entailment Model Checking}

\author{Tatpong Katanyukul} 
 \email[Corresponding author: ]{tatpong@kku.ac.th}
\affiliation{
Computer Engineering, Khon Kaen University\\
Khon Kaen, Thailand 40002.
}


\date{\today} 

\begin{abstract}
Targeting entailment model checking, 
a recent study has pioneered an idea of Eigenmarking search, an improvement over Grover search using extra qubits.
The extra qubits condition the quantum state evolution such that
the answer states (if exist) are always in the minority.
The minority criteria is essential to Grover probability-amplitude amplification
and consequently the effectiveness of Grover search.
In addition to enforce the minority criteria, 
Eigenmarking also employs complementary states (through well-orchestrated phase rotation) 
for easy identification of a no-answer case (related to a no-violation case in the context of model checking).
Eigenmarking search has been shown effective in two-qubit simulations.

The three Eigenmarking schemes have been previously proposed.
Two schemes require two extra qubits.
One scheme (called ``subtle marking'') requires one extra qubit with a multiple-qubit-controlled phase rotation.
Our study refines the mechanism using only one extra qubit with only two-qubit-controlled phase rotation, commonly known as \texttt{ccz},
regardless of how many qubits the input has.
Using a multiple-qubit-controlled phase rotation (as in subtle marking) 
associates with highly entangled states. 
Highly entangled states in a real quantum hardware are difficult
(or in some cases may even be unachievable)
particularly in a scaled up scenario involving many qubits.
Our proposed new Eigenmarking scheme has lightened the burden for the hardware requirement.
The new Eigenmarking search has been experimented in two-qubit-system simulations
and shown viable, achieving the minimal relative local winning margin of $W=3.17$
and the worst-case distinguishability of $D=0.769$ (cf. $W=0.67$; $D=0.19$ from conventional marking and $W=0.28$; $D=0.55$ from subtle marking).

\end{abstract}

\maketitle

\section{\label{sec:intro}Introduction}

Logical reasoning is one of the most notable functions
in computational intelligence.
Entailment --- truth inference of an under-questioned sentence
based on an accepted knowledge ---
is central to resolve logical reasoning.
Entailment
can be verified using model checking:
checking every combination of truth values for all logical symbols involved (the input)
and their resultant truth values if the resultant values of the sentence agrees with ones of the knowledge.
However, 
computation cost of model checking 
may be exponential of 
the number of possible input values\cite{RN2022}.

The advent of rapid development in quantum computing\cite{KimEtAl2023}
has offered a great opportunity 
for computational intensive tasks.
A previous study\cite{eigen2025}
has proposed Eigenmarking schemes
to enhance Grover search\cite{Grover97} 
in the context of entailment model checking.
Original Grover search assumes a single-answer scenario
and exploit the minority condition in one of its key mechanism, i.e.,
probability-amplitude amplification.

The context of entailment model checking
requires a huge relaxation on this minority constraint.
The key idea of Eigenmarking
is to add extra qubits such that 
regardless of a proportion of answer states in all possible input states,
these answer states remain minorities in a total number of all states.
Another word, extra qubits create many dummy states such that even all input states are answers,
they still remain minorities in this extended state space:
one extra qubit allows double the number of states;
two extra qubits allow quadruple the number of states.

While the extra qubits have extended the state space,
the extra qubits also act as a tag for the input states:
the first previously proposed scheme\cite{eigen2025} 
(called ``conventional marking'' from now on) 
uses tag \texttt{01} to mark the answer states (target group)
and tag \texttt{10} for the complementary states; 
the last previously proposed scheme\cite{eigen2025} (called ``subtle marking'') 
uses tag \texttt{0} for the answers
and tag \texttt{1} for the complementaries.

In addition to imposing the minority condition,
to efficiently handle a no-answer case
some complementary states 
are selected to 
represent a no-answer scenario:
the probability amplitudes of no-answer-representive states
are considerably amplified when there is no answer.

Previous Eigenmarking schemes have achieved such functionalities using:
two extra qubits and $\pi/2$-phase rotations in one of the schemes 
(we will call this scheme ``conventional marking'')
and
one extra qubit and multiple-qubit-controlled phase rotation in another 
(called ``subtle marking'').

Conventional marking is reported to have 
a good global winning margin, i.e.,
the answer states have remarkably larger chances to be measured (regardless of a tag value),
but when there is no answer, 
it has to rely on the difference between measurements in the target states 
and the ones in complementary states to decide if it actually is a no-answer scenario.

Subtle marking is reported to have
a good local winning margin ---
when considered among the target states (with target tag, i.e., tag qubit is \texttt{0})
the answer states have also substantially larger chances to be measured ---
and a large contrast between measurements in the target states (having very low chance)
and ones in complementary states (having very high chance) when there is no answer.
Its second reported property 
makes determining if there is any answer easier 
when using subtle marking.

Although subtle marking is reported to have bad global winning margin
---
there are large chances that non-answer states with a non-target tag  
may be observed ---
, this may be easily mitigated by discarding any state with a non-target tag.
However a main disadvantage of subtle marking is architectural:
it uses multiple-qubit-controlled phase rotation.
This puts too much stress on the hardware requirement
and for a large-input problem, e.g., $1000$-qubit input,
this may even be impossible.

Our work proposed a more refined scheme to reduce stress on hardware requirement
and evaluate the quality of its deliverables conferring to 
conventional and subtle Eigenmarking schemes.

\section{\label{sec:background}Background}

\paragraph{Entailment.}
When an accepted knowldge $\alpha$ entails 
a sentence $\beta$,
it means that 
what sentence $\beta$ says
agrees with what knowldge $\alpha$ has said.
This is written with notation $\alpha \vDash \beta$.
It means that 
for every combination of input truth values
when the entailer $\alpha$ is true,
the entailee $\beta$ must be true.

There are two main approaches for entailment checking\cite{RN2022}:
\emph{theorem proving}
and \emph{model checking}.
Theorem proving applies rules of inference to the entailer 
to eventually derive the entailee.
Model checking checkes all input truth values
to validate the resultant truth values.

E.g., 
suppose knowldge $\alpha = \{ A \Rightarrow B, B \Rightarrow C \}$ ( ``Monthong is a variety of durian.'', ``Durian is prickly.'');
$\beta_1 = \{ A \Rightarrow C \}$ (``Monthong is prickly.'');
$\beta_2 = \{ A \Rightarrow \neg C \}$ (``Monthong is not prickly.'');
$\beta_3 = \{ D \Rightarrow E \}$ (``There is life on Mars.''); and
$\beta_4 = \{ D \Rightarrow \neg E \}$ (``There is no life on Mars.''),
this can be reasoned:
$\alpha \vDash \beta_1$;
$\alpha \nvDash \beta_2$;
$\alpha \nvDash \beta_3$ and
$\alpha \nvDash \beta_4$
for violations as shown in Table~\ref{tab:modelcheck} (violation is emphasized in bold).
For entailment model checking, 
entailment is equivalent to no violation: 
for all $x$'s
no logical false of $\neg \alpha(x) \vee \beta(x)$
($\equiv$ no logical true of $\alpha(x) \wedge \neg \beta(x)$).

\begin{table}
\caption{\label{tab:modelcheck}Example of entailment model checking: $\alpha \vDash \beta$ if and only if $\neg \alpha(x) \vee \beta(x)$ for all $x$'s.}
\begin{ruledtabular}
\begin{tabular}{ccccccccccl}
A & B & C & D & E & $\alpha$ & $\beta_1$ & $\beta_2$ & $\beta_3$ & $\beta_4$ & Note \\
 \hline
F & F & F & F& F &   T       & T         & T         & T         & T         &      \\

$\vdots$ & $\vdots$ & $\vdots$ & $\vdots$ & $\vdots$ &   $\vdots$       & $\vdots$         & $\vdots$         & $\vdots$         & $\vdots$         &      \\

T & T & T & F & F &   T       & T         & \textbf{F}         & T         & T         &  \\
T & T & T & F & T &   T       & T         & \textbf{F}         & T         & T         & \\
T & T & T & T & F &   T       & T         & \textbf{F}         & \textbf{F}         & T         &  \\
T & T & T & T & T &   T       & T         & \textbf{F}         & T         & \textbf{F}         &  \\
\end{tabular}
\end{ruledtabular}
\end{table}

\paragraph{Quantum computation.}
Quantum properties, such as superposition and entanglement, can be exploited for computation.
Superposition is a quantum state 
that is a linear combination of multiple eigenstates,
e.g., state $| \psi \rangle = a_i | + \rangle + b_i |-\rangle$
where $| + \rangle$ and  $|-\rangle$ are the eigenstates (depending on the measurement operator)
and $a_i$ and $b_i$ are probability amplitudes of the corresponding eigentstates.
For a given basis (a set of eigenstates), 
a state can also be written in a vector form 
$| \psi \rangle = [a_i , b_i]^T$,
where
the eigenstates $| + \rangle$ and $| - \rangle$ are implicit.

A state can evolve in two distinct modes\cite{NC2016}.
(1) In quantum evolution mode, 
state $| \psi \rangle$ evolves according to Schr{\"o}dinger equation (SE)
based on Hamiltonian $\hat{H}$ (energy operator) of the system.
I.e.,
$i \hbar \frac{\partial | \psi \rangle}{\partial t} =  \hat{H} | \psi \rangle$
where $i = \sqrt{-1}$ and $\hbar$ is a reduced Planck constant.
The solution to SE,
$| \psi \rangle = | \psi_0 \rangle \exp(-\frac{i}{\hbar} \hat{H} t)$
where $| \psi_0 \rangle$ is an initial state.

For a specific period of time $t$, 
this is equivalent to \textit{unitary transformation}
whose unitary operator corresponds to the Hamiltonian:
$U \equiv \exp(-\frac{i}{\hbar} \hat{H} t)$, hence
$| \psi \rangle = U | \psi_0 \rangle$
or a state is a unitary transformation from its previous state: 
$| \psi' \rangle = U | \psi \rangle$.
A control on system energy
dictates the Hamiltonian 
and equivalently manipulates a unitary operator.
Some common unitary operators are 
\texttt{not} operator 
$X = \begin{bmatrix}0 & 1 \\ 1 & 0\end{bmatrix}$,
Hadamard operator 
$H = \frac{1}{\sqrt{2}}\begin{bmatrix}1 & 1\\1 & -1\end{bmatrix}$,
and one-qubit-controlled-phase 
$CZ = \begin{bmatrix}
      1 & 0 & 0 & 0 \\
      0 & 1 & 0 & 0 \\
      0 & 0 & 1 & 0 \\
      0 & 0 & 0 & -1 
      \end{bmatrix}$ for a two-qubit system.

(2) In measurement mode, 
state collapses to one of its eigenstates upon measurement.
The eigenstates correspond to the measurement operator, 
which is applied to measure the system.
A probability of an eigenstate to which the state collapses
is a squared modulus of the amplitude of that eigenstate in the state before the measurement.
E.g., 
given a state $| \psi \rangle = a_i | + \rangle + b_i |-\rangle = [a_i, b_i]^T$ prior to the measurement in
a basic of $| + \rangle$ and $|-\rangle$,
after the measurement
the state will collapse to either $| + \rangle$ (with probability $|a_i|^2$),
i.e., $| \psi' \rangle = | + \rangle = [1, 0]^T$,
or $| - \rangle$ (with probability $|b_i|^2$),
i.e., $| \psi' \rangle = | - \rangle = [0, 1]^T$.

A hardware of a quantum computer can be implemented 
using many technologies, e.g., superconducting circuit\cite{KKYOGO19},
and to control qubits can be done by changing the system energy, 
e.g., through microwave pulses.

\begin{figure}[htbp]
    \begin{tabular}{cccc}
    \includegraphics[height=2.5cm]{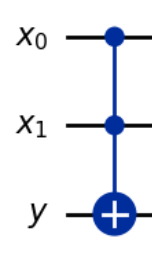}
    &
    \includegraphics[height=2.5cm]{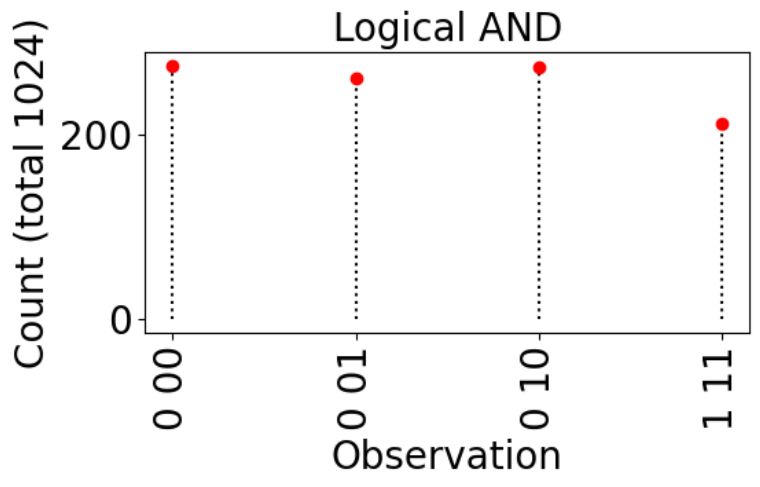}
    &
    \includegraphics[height=2.5cm]{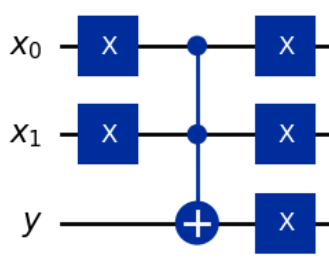}
    &
    \includegraphics[height=2.5cm]{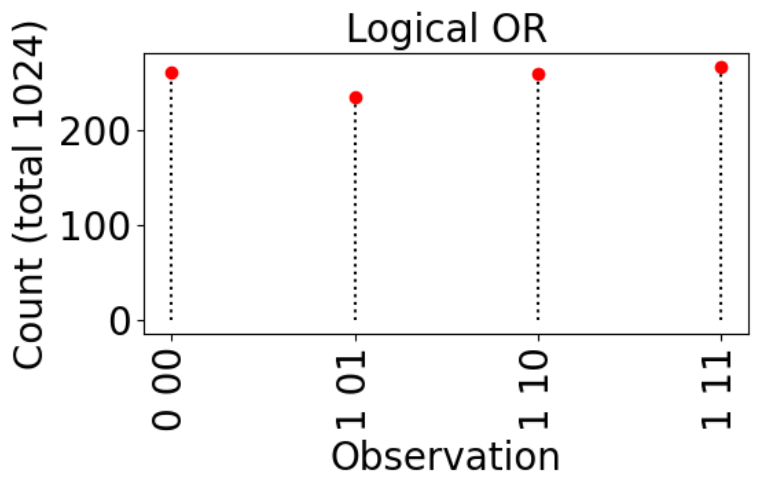}
    \\
    (a) & (b) & (c) & (d)
    \end{tabular}
\caption{Quantum gate applications: (a) logical AND, (b) its simulated result, (c) logical OR, and (d) its simulated result.}
\label{fig: classical logic}
\end{figure}

\paragraph{Classical logic from quantum gates.}
Any classical logic can be derived from a proper application of quantum gates.
Figure~\ref{fig: classical logic} shows equivalent functions to logical AND and OR.

\paragraph{Grover search.}
Grover algorithm\cite{Grover97} addresses a search problem:
given a function $f: \{0, 1\}^n \rightarrow \{0, 1\}$
with a promise that
there is exactly one answer $\mathbf{x}'$ 
such that
$f(\mathbf{x}) = 1$ if $\mathbf{x} = \mathbf{x}'$ and $f(\mathbf{x}) = 0$ otherwise.
A binary input $\mathbf{x}$ has $n$ qubits.
Thus, there are $N = 2^n$ possible combinations to search.

Suppose 
the underlying function is given as a corresponding unitary operator $U_f$, 
which
takes $n + 1$ qubits ($n$ for input $\mathbf{x}$ and $1$ for ancillary $y$)
and rotates $y$ qubit by $\pi$ (phase inversion) only when $f(\mathbf{x}) = 1$.
That is, the unknown function operator
$U_f \equiv |\mathbf{x}'\rangle \langle \mathbf{x}'| \otimes R_z(\pi)  + \sum_{\mathbf{x} \in \tilde{X}} |\mathbf{x}\rangle \langle \mathbf{x}| \otimes I$
where $\tilde{X}$ is a set of non-winning states;
identity $I = \begin{bmatrix} 1 & 0 \\ 0 & 1\end{bmatrix}$
and rotation $R_z(\theta) = \begin{bmatrix} 1 & 0 \\ 0 & e^{i \theta}\end{bmatrix}$.


Grover algorithm 
is as follows:
(1) Prepare $|\mathbf{x} \rangle$ in a ground state, 
          i.e., $|\mathbf{x}_0 \rangle = |\mathbf{0}\rangle \equiv |\underbrace{00\ldots0}_n \rangle$. 
(2) Apply $H^{\otimes n}$,
            where $H^{\otimes n} = \underbrace{H \otimes H \otimes \cdots \otimes H}_{n}$ 
            is a tensor product of $H$.
            It is equivalent to apply $H$ independently to each qubit of $| \mathbf{x}_0 \rangle$,
            i.e.,
            $|\mathbf{x}_1\rangle = H^{\otimes n} |\mathbf{x}_0\rangle$.
(3) Apply the phase inversion, 
            i.e., 
            $|\psi_2\rangle = U_f |\mathbf{x}_1, 1 \rangle$.
(4) Apply the inversion about the mean to the input part,
            i.e.,
            segment $|\mathbf{x}_2 \rangle |y \rangle = | \psi_2 \rangle$
            and have
            $|\mathbf{x}_3 \rangle = (2 A - I) |\mathbf{x}_2 \rangle$,
            where $I$ is an identity matrix and an average matrix 
            $A = \frac{1}{N} \begin{bmatrix}
            1 & 1 & \ldots & 1 \\
            1 & 1 & \ldots & 1\\
            \vdots & \vdots & \ddots & \vdots \\
            1 & 1 & \ldots & 1
            \end{bmatrix}$.
(5) Grover operation (steps 3 and 4) should be applied for $J$ times (Derived from \cite{BBHT96}),
$J = \arg\min_j \frac{1}{N - 1} \cos^2 \left( (2j + 1) \sin^{-1}\sqrt{\frac{1}{N}} \right)$
or simply $J = \mathrm{round}\left(\frac{\pi}{4} \sqrt{N} - \frac{1}{2}\right)$ when $N$ is large.
(6) Measure the qubits.
An eigenstate observed in the measurement is very likely to be the answer.

\paragraph{Eigenmarking.}
Conventional Eigenmarking\cite{eigen2025} (1) adds two additional qubits 
$|t_1 t_0\rangle$ 
each controls phase rotation on the ancillary $|y\rangle$
and 
(2) has $U_f$ shift phase by only $\pi/2$ 
(instead of full phase inversion, equivalent to $\pi$),
and 
(3) extends 
the inversion about the mean 
to average over all qubits (not just the input qubits).
Table~\ref{tab: conventional marking} shows the algorithm.

\begin{table}
    \caption{\label{tab: conventional marking}Conventional marking.}
    \begin{tabular}{l}
        (1) Have
        $U_f \equiv |\mathbf{x}'\rangle \langle \mathbf{x}'| \otimes R_z(\pi/2) + \sum_{\mathbf{x} \in \tilde{X}} |\mathbf{x}\rangle \langle \mathbf{x}| \otimes I$.
        \\
        (2) Prepare
        $|\psi_0\rangle = |\mathbf{x}, y \rangle = |\mathbf{0},0\rangle$.
        \\
        (3) Do
        $|\psi_1\rangle = H^{\otimes (n+1)} |\psi_0\rangle$.
        \\
        (4) Apply Grover selection $|\psi_2\rangle = U_f |\psi_1\rangle$.
        \\
        (5) Apply marking:
        let $|t_1, t_0\rangle = H^{\otimes 2} |0,0\rangle$
        and segment $|\mathbf{x}_2\rangle | y_2 \rangle = |\psi_2\rangle$,
        then $|t_0', y_3 \rangle = \mathbf{CR}(\pi/2) |t_0, y_2 \rangle$
        \\
        and $|t_1', y_4 \rangle = \mathbf{CR}(-\pi/2) |t_1, y_3 \rangle$
        where $\mathbf{CR}(\theta) = I \otimes |0\rangle \langle 0| + R_z(\theta) \otimes |1\rangle \langle 1|$.
        \\
        (6) Apply inversion about the mean to all qubits:
        aggregate $| \psi_4 \rangle \equiv |t_1', t_0', \mathbf{x}_2, y_4 \rangle$
        and $| \psi_5 \rangle = (2 A - I) | \psi_4 \rangle$.
        \\
        (7) Measure the qubits.        
    \end{tabular}
    \end{table}



Subtle marking\cite{eigen2025} employs only one extra qubit, 
but resorts to multiple-qubit-controlled phase rotation.
Table~\ref{tab: subtle mark} shows subtle marking algorithm. 

\begin{table}
    \caption{\label{tab: subtle mark}Subtle marking.}
    \begin{tabular}{l}
(1) Prepare system in the ground state: $|\psi_0 \rangle = | \mathbf{0} \rangle$,
where $|\psi_0 \rangle = |t_0, \mathbf{x}, y \rangle$.
\\
(2) Apply Hadamard: $|\psi_1 \rangle = H^{\otimes (n+2)} |\psi_0 \rangle$.
\\
(3) Apply Grover selection: $|\psi_2 \rangle = (I \otimes U_f) |\psi_1 \rangle$.
\\
(4) Apply marking: $|\psi_3 \rangle = \mathrm{MCR}(\pi)|\psi_2 \rangle$,
where $\mathrm{MCR}(\theta) = | 1,\mathbf{1} \rangle \langle 1,\mathbf{1}| \otimes R_z(\theta)  + \sum_{\mathbf{x} \neq | 1, \mathbf{1} \rangle} | \mathbf{x} \rangle \langle \mathbf{x}| \otimes I$.
\\
(5) Do inversion about the mean: $|\psi_4 \rangle = (2 A - I) |\psi_3 \rangle$.
\\
(6) Measure the qubits.
\end{tabular}
\end{table}


\section{\label{sec:simplereigen}Simpler Eigenmarking}
Our work refines subtle marking to use only two-qubit controlled phase rotation, 
significantly reduce stress on the hardware requirement.
Specifically, step (4) in Table~\ref{tab: subtle mark}
is changed to:
\begin{itemize}
    \item (4.1) Separate out the controlling and controlled qubits: $|\hat{\mathbf{x}} \rangle, |t, x_{\mbox{msq}}, y \rangle \equiv | \psi_2 \rangle$ 
    where $|\hat{\mathbf{x}}\rangle$ is the state of $n-1$ least significant input qubits;
    $t$ is a tag qubit;
    $x_{\mbox{msq}}$ is the most significant qubit of the input;
    $y$ is an ancillary;
    \item (4.2) Apply marking operation: $|t',x'_{\mbox{msq}},y' \rangle = \mathrm{CCZ}|t,x_{\mbox{msq}},y \rangle$; and
    \item (4.3) Assemble qubits back: $| \psi_3 \rangle = |\hat{\mathbf{x}} \rangle, |t', x'_{\mbox{msq}}, y' \rangle$.
\end{itemize}

The operator $\mathrm{CCZ}$ is a two-qubit controlled phase rotation:
\begin{align}
\mathrm{CCZ} &= 
| 1,1 \rangle \langle 1,1| \otimes R_z(\pi)  
+ \sum_{\mathbf{x} \neq | 1, 1 \rangle} | \mathbf{x} \rangle \langle \mathbf{x}| \otimes I
=
%
\begin{bmatrix}
                 1 & 0 & 0 & 0 & 0 & 0 & 0 & 0 \\
                 0 & 1 & 0 & 0 & 0 & 0 & 0 & 0 \\
                 0 & 0 & 1 & 0 & 0 & 0 & 0 & 0 \\
                 0 & 0 & 0 & 1 & 0 & 0 & 0 & 0 \\
                 0 & 0 & 0 & 0 & 1 & 0 & 0 & 0 \\
                 0 & 0 & 0 & 0 & 0 & 1 & 0 & 0 \\
                 0 & 0 & 0 & 0 & 0 & 0 & 1 & 0 \\
                 0 & 0 & 0 & 0 & 0 & 0 & 0 & -1 \\
                \end{bmatrix}
\nonumber .
\end{align}

Since the ancillary $y$ is put through Hadamard gate and it is eventually averaged over,
this qubit also acts as an extra tag (and plays a role in extending a state space).
Therefore, this qubit is also measured at the end of the algorithm (step 6).

\begin{figure}[htbp]
    \begin{tabular}{cc}
    \includegraphics[height=3cm]{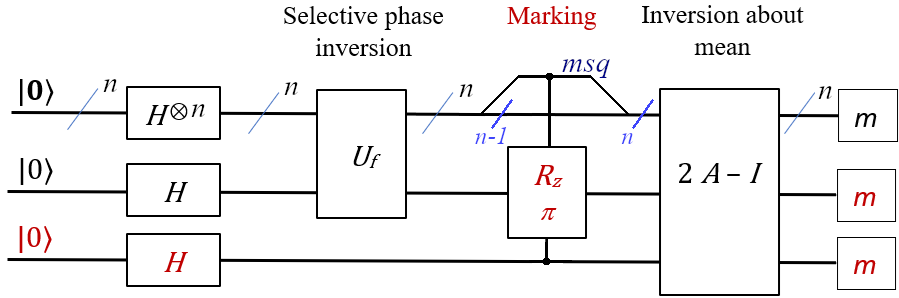}
    &
    \includegraphics[height=4cm]{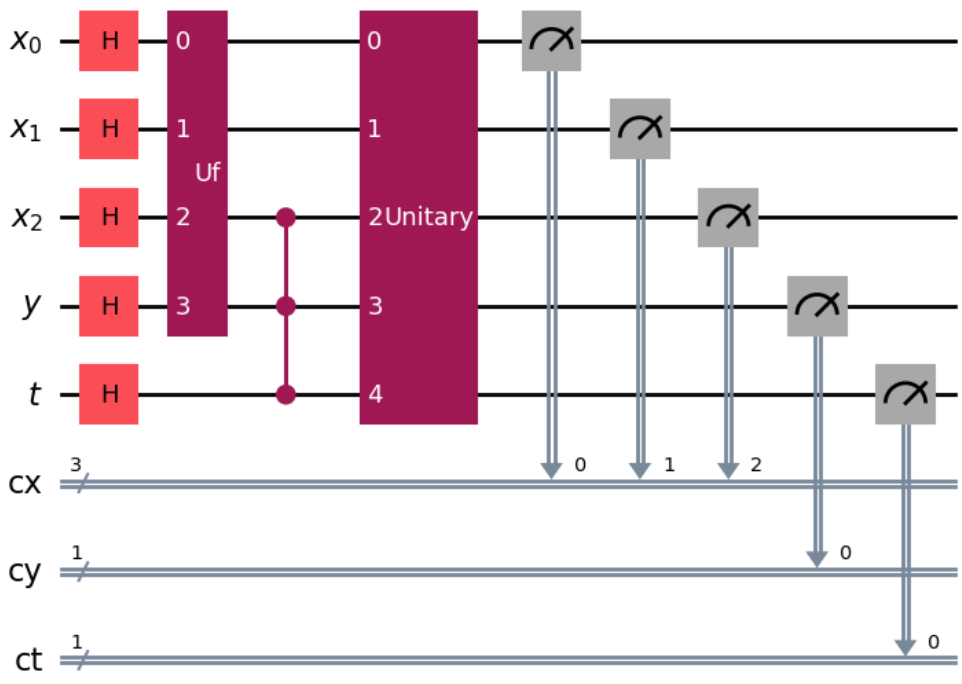}
    \end{tabular}
\caption{Simpler Eigenmarking. Generic diagram (left) and 
         Qiskit graphic draw of a three-qubit system (right): $x_2$ is the most significant qubit (msq).}
\label{fig: simpler mark}
\end{figure}

\section{\label{sec:methodology}Methodology}

Our proposed new Eigenmarking scheme is tested on all possible scenarios 
of a two-qubit system
using Qiskit (version 1.1.1) and Qiskit Aer Simulator (version 0.15.1).
Each treatment is repeated for 40 times, 
while each time the simulator simulates it for 1024 runs.
The winning margins $W$ 
and distinguishabilities $D$
are presented in
Table~\ref{tab: win margins}.
Both performance indices are introduced in the previous study\cite{eigen2025}.
The winning margin is a  relative difference between 
the minimal count of the answer states 
and the maximal count of the non-answer states.
Distinguishability 
quantifies
the worst-case difference 
between the minimal score in scenarios with at least some answer(s) 
and the maximal score in a no-answer scenario:
$D = \min \{\min(M_i)\}_{i > 0} - \max(M_0)$
where $M_i$ is a marking factor of the scenario with $i$-answer(s)
and $M_i = \frac{c_t - c_n}{c_t + c_n}$; 
$c_t$ is a maximum count of states with target tag (simpler marking: $ty=$ \verb|01|),
$c_n$ is a maximum count of states representing a no-answer case
(simpler marking: $tyx_{msq}=$ \verb|111|).
A large value of $D$ indicates the ease of differentiate a no-answer case from a some-answer one.


\begin{table}[htbp]
    \caption{Winning margin [min., mean, max.](std. dev.) and distinguishability}
    \begin{center}
    \begin{tabular}{|l|c|c|c|}
    \hline
    \textbf{Scheme} &  \multicolumn{2}{|c|}{\textbf{Relative winning margin}} & \textbf{Distinguishability} \\
    \cline{2-3}
                    & \textbf{Global}   & \textbf{Local}                      &  $D$                        \\
    \hline
    Convention.  & [0.57,  \textbf{1.10}, 1.76](0.2)  & [0.67, \textbf{1.49}, 2.60](0.3)     & 0.190  \\
    Subtle.      & [-0.37, \textbf{0.31}, 6.12](1.4)  & [0.28, \textbf{25.72}, 197.00](15.8) & 0.550 \\
    Simpler.     & [-0.29, \textbf{1.06}, 19.95](4.1) & [3.17, \textbf{Large}, \textbf{Large}](\textbf{Large})\footnote{Count of any non-answer state with a target tag is zero.} & 0.769 \\
    \hline
    \end{tabular}
    \label{tab: win margins}
    \end{center}
\end{table}


\section{\label{sec:conclusion}Conclusion and Discussion}

A new Eigenmarking scheme is proposed.
It requires only one extra qubit and a common \verb|ccz| gate.
Its performance has been shown to be viable in a two-qubit simulation.
However, to further advance the development, 
its scalability (scaling up to a large number of qubits),
robustness (testing on a real quantum machine),
and theoretical analysis
have to be adequately investigated. 


\bibliography{eigen}

\providecommand{\noopsort}[1]{}\providecommand{\singleletter}[1]{#1}%
\begin{thebibliography}{7}%
\makeatletter
\providecommand \@ifxundefined [1]{%
 \@ifx{#1\undefined}
}%
\providecommand \@ifnum [1]{%
 \ifnum #1\expandafter \@firstoftwo
 \else \expandafter \@secondoftwo
 \fi
}%
\providecommand \@ifx [1]{%
 \ifx #1\expandafter \@firstoftwo
 \else \expandafter \@secondoftwo
 \fi
}%
\providecommand \natexlab [1]{#1}%
\providecommand \enquote  [1]{``#1''}%
\providecommand \bibnamefont  [1]{#1}%
\providecommand \bibfnamefont [1]{#1}%
\providecommand \citenamefont [1]{#1}%
\providecommand \href@noop [0]{\@secondoftwo}%
\providecommand \href [0]{\begingroup \@sanitize@url \@href}%
\providecommand \@href[1]{\@@startlink{#1}\@@href}%
\providecommand \@@href[1]{\endgroup#1\@@endlink}%
\providecommand \@sanitize@url [0]{\catcode `\\12\catcode `\$12\catcode
  `\&12\catcode `\#12\catcode `\^12\catcode `\_12\catcode `\%12\relax}%
\providecommand \@@startlink[1]{}%
\providecommand \@@endlink[0]{}%
\providecommand \url  [0]{\begingroup\@sanitize@url \@url }%
\providecommand \@url [1]{\endgroup\@href {#1}{\urlprefix }}%
\providecommand \urlprefix  [0]{URL }%
\providecommand \Eprint [0]{\href }%
\providecommand \doibase [0]{http://dx.doi.org/}%
\providecommand \selectlanguage [0]{\@gobble}%
\providecommand \bibinfo  [0]{\@secondoftwo}%
\providecommand \bibfield  [0]{\@secondoftwo}%
\providecommand \translation [1]{[#1]}%
\providecommand \BibitemOpen [0]{}%
\providecommand \bibitemStop [0]{}%
\providecommand \bibitemNoStop [0]{.\EOS\space}%
\providecommand \EOS [0]{\spacefactor3000\relax}%
\providecommand \BibitemShut  [1]{\csname bibitem#1\endcsname}%
\let\auto@bib@innerbib\@empty
\bibitem [{\citenamefont {Russell}\ and\ \citenamefont
  {Norvig}(2022)}]{RN2022}%
  \BibitemOpen
  \bibfield  {author} {\bibinfo {author} {\bibfnamefont {S.}~\bibnamefont
  {Russell}}\ and\ \bibinfo {author} {\bibfnamefont {P.}~\bibnamefont
  {Norvig}},\ }\href@noop {} {\emph {\bibinfo {title} {Artificial intelligence:
  a modern approach}}}\ (\bibinfo  {publisher} {Pearson},\ \bibinfo {year}
  {2022})\BibitemShut {NoStop}%
\bibitem [{\citenamefont {Kim}\ \emph {et~al.}(2023)\citenamefont {Kim},
  \citenamefont {Eddins}, \citenamefont {Anand},\ and\ \citenamefont
  {et~al.}}]{KimEtAl2023}%
  \BibitemOpen
  \bibfield  {author} {\bibinfo {author} {\bibfnamefont {Y.}~\bibnamefont
  {Kim}}, \bibinfo {author} {\bibfnamefont {A.}~\bibnamefont {Eddins}},
  \bibinfo {author} {\bibfnamefont {S.}~\bibnamefont {Anand}}, \ and\ \bibinfo
  {author} {\bibnamefont {et~al.}},\ }\bibfield  {title} {\enquote {\bibinfo
  {title} {Evidence for the utility of quantum computing before fault
  tolerance},}\ }\href@noop {} {\bibfield  {journal} {\bibinfo  {journal}
  {Nature}\ }\textbf {\bibinfo {volume} {618}},\ \bibinfo {pages} {500--505}
  (\bibinfo {year} {2023})}\BibitemShut {NoStop}%
\bibitem [{\citenamefont {Katanyukul}(2025)}]{eigen2025}%
  \BibitemOpen
  \bibfield  {author} {\bibinfo {author} {\bibfnamefont {T.}~\bibnamefont
  {Katanyukul}},\ }\bibfield  {title} {\enquote {\bibinfo {title} {Toward
  entailment checking: explore eigenmarking search},}\ }\href@noop {}
  {\bibfield  {journal} {\bibinfo  {journal} {in InCACCT}\ } (\bibinfo {year}
  {2025})}\BibitemShut {NoStop}%
\bibitem [{\citenamefont {Grover}(1997)}]{Grover97}%
  \BibitemOpen
  \bibfield  {author} {\bibinfo {author} {\bibfnamefont {L.}~\bibnamefont
  {Grover}},\ }\bibfield  {title} {\enquote {\bibinfo {title} {Quantum
  mechanics helps in searching for a needle in a haystack},}\ }\href@noop {}
  {\bibfield  {journal} {\bibinfo  {journal} {Phys. Rev. Lett.}\ }\textbf
  {\bibinfo {volume} {79(325)}} (\bibinfo {year} {1997})}\BibitemShut {NoStop}%
\bibitem [{\citenamefont {Nielsen}\ and\ \citenamefont
  {Chuang}(2016)}]{NC2016}%
  \BibitemOpen
  \bibfield  {author} {\bibinfo {author} {\bibfnamefont {M.~A.}\ \bibnamefont
  {Nielsen}}\ and\ \bibinfo {author} {\bibfnamefont {I.~L.}\ \bibnamefont
  {Chuang}},\ }\href@noop {} {\emph {\bibinfo {title} {Quantum computation and
  quantum information}}}\ (\bibinfo  {publisher} {CUP},\ \bibinfo {year}
  {2016})\BibitemShut {NoStop}%
\bibitem [{\citenamefont {Krantz}\ \emph {et~al.}(2019)\citenamefont {Krantz},
  \citenamefont {Kjaergaard}, \citenamefont {Yan},\ and\ \citenamefont
  {et~al.}}]{KKYOGO19}%
  \BibitemOpen
  \bibfield  {author} {\bibinfo {author} {\bibfnamefont {P.}~\bibnamefont
  {Krantz}}, \bibinfo {author} {\bibfnamefont {M.}~\bibnamefont {Kjaergaard}},
  \bibinfo {author} {\bibfnamefont {F.}~\bibnamefont {Yan}}, \ and\ \bibinfo
  {author} {\bibnamefont {et~al.}},\ }\bibfield  {title} {\enquote {\bibinfo
  {title} {A quantum engineer's guide to superconducting qubits},}\ }\href@noop
  {} {\bibfield  {journal} {\bibinfo  {journal} {Appl. Phys. Rev.}\ }\textbf
  {\bibinfo {volume} {6(021318)}} (\bibinfo {year} {2019})}\BibitemShut
  {NoStop}%
\bibitem [{\citenamefont {Boyer}\ \emph {et~al.}(1998)\citenamefont {Boyer},
  \citenamefont {Brassard}, \citenamefont {H{\o}yer},\ and\ \citenamefont
  {Tapp}}]{BBHT96}%
  \BibitemOpen
  \bibfield  {author} {\bibinfo {author} {\bibfnamefont {M.}~\bibnamefont
  {Boyer}}, \bibinfo {author} {\bibfnamefont {G.}~\bibnamefont {Brassard}},
  \bibinfo {author} {\bibfnamefont {P.}~\bibnamefont {H{\o}yer}}, \ and\
  \bibinfo {author} {\bibfnamefont {A.}~\bibnamefont {Tapp}},\ }\bibfield
  {title} {\enquote {\bibinfo {title} {Tight bounds on quantum search},}\
  }\href@noop {} {\bibfield  {journal} {\bibinfo  {journal} {Fortschritte der
  Physik}\ }\textbf {\bibinfo {volume} {46}} (\bibinfo {year}
  {1998})}\BibitemShut {NoStop}%
\end{thebibliography}%

\section{Appendix}

\subsection{Illustrative plots}

\begin{figure}[htbp]
\begin{tabular}{cccc}
\includegraphics[width=0.24\textwidth]{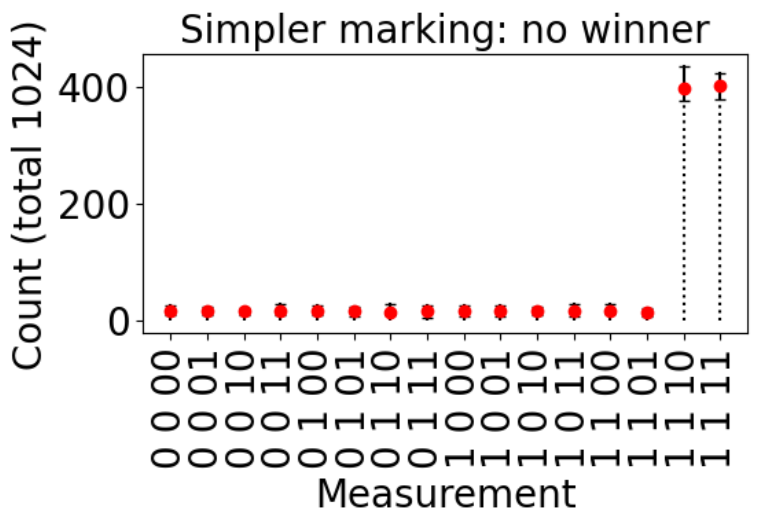}
&
\includegraphics[width=0.24\textwidth]{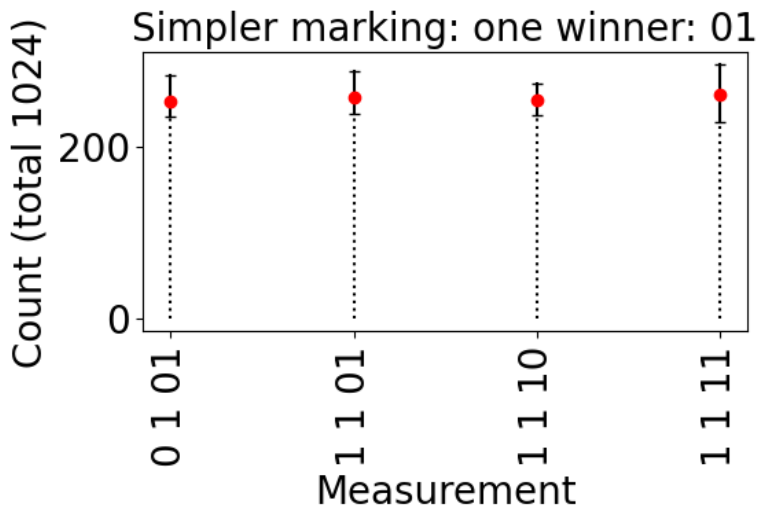}
&
\includegraphics[width=0.24\textwidth]{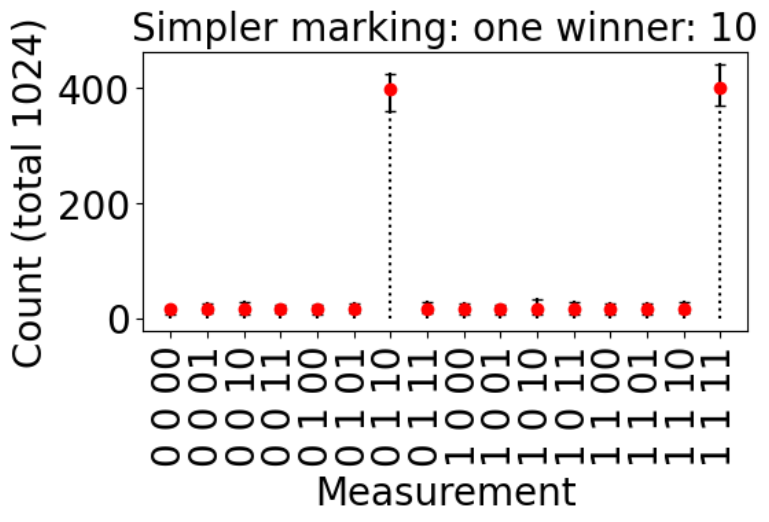}
&
\includegraphics[width=0.24\textwidth]{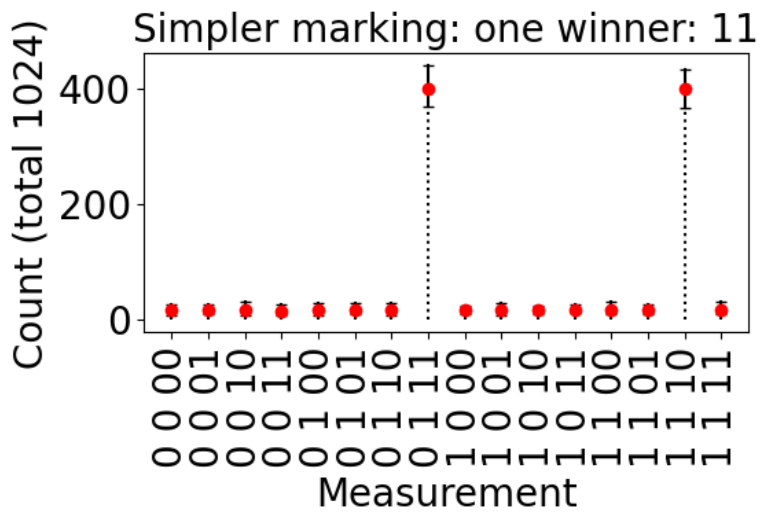}
\\
\includegraphics[width=0.24\textwidth]{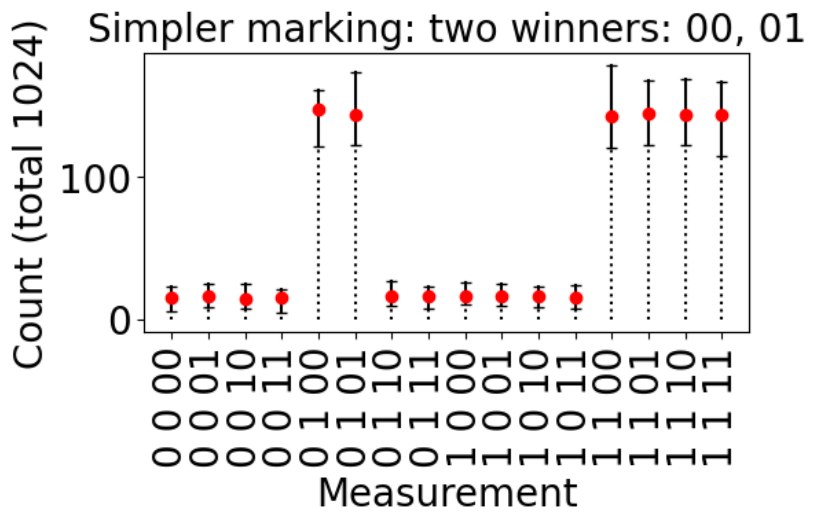}
&
\includegraphics[width=0.24\textwidth]{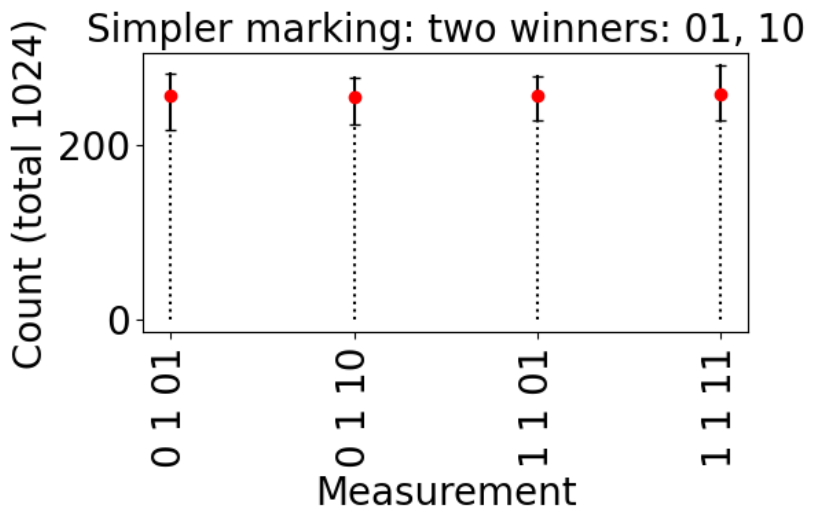}
&
\includegraphics[width=0.24\textwidth]{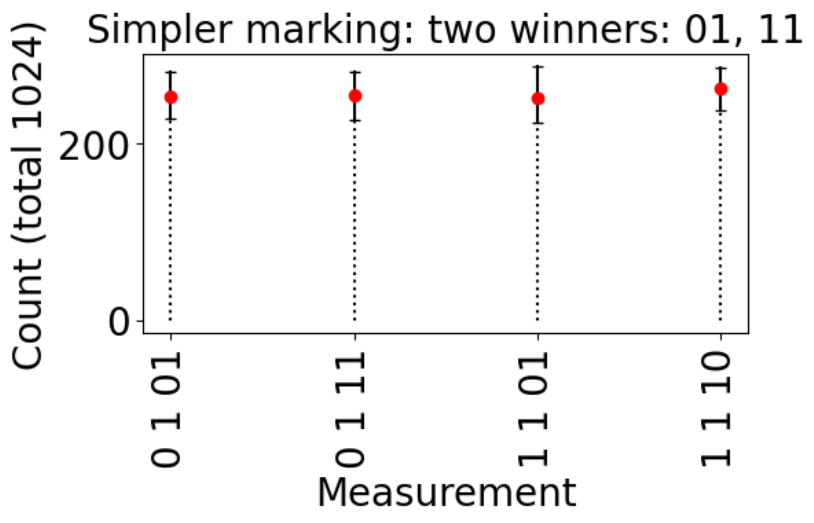}
&
\includegraphics[width=0.24\textwidth]{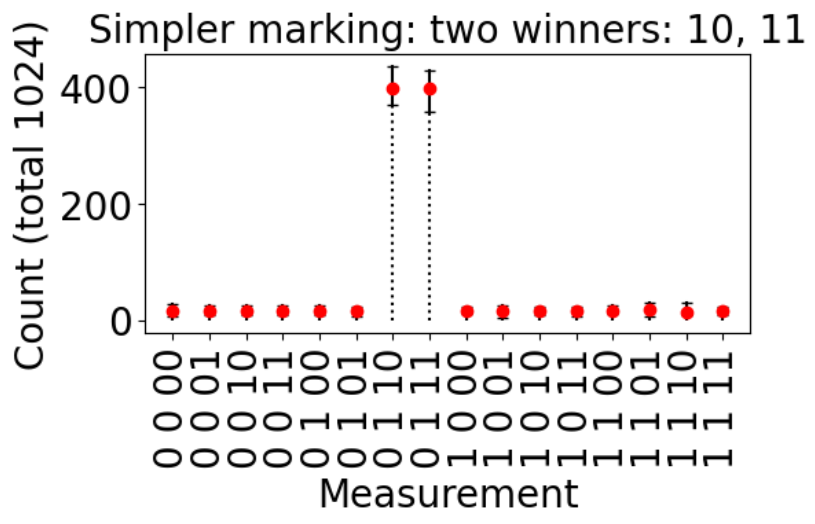}
\\

\includegraphics[width=0.24\textwidth]{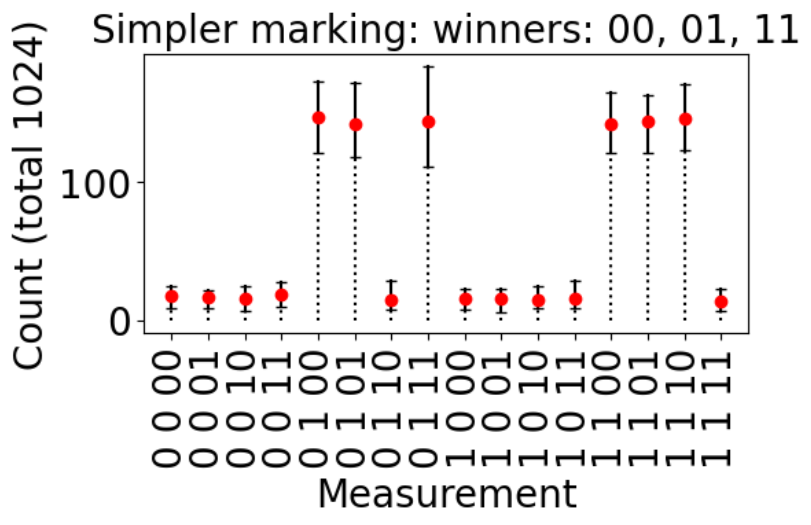}
&
\includegraphics[width=0.24\textwidth]{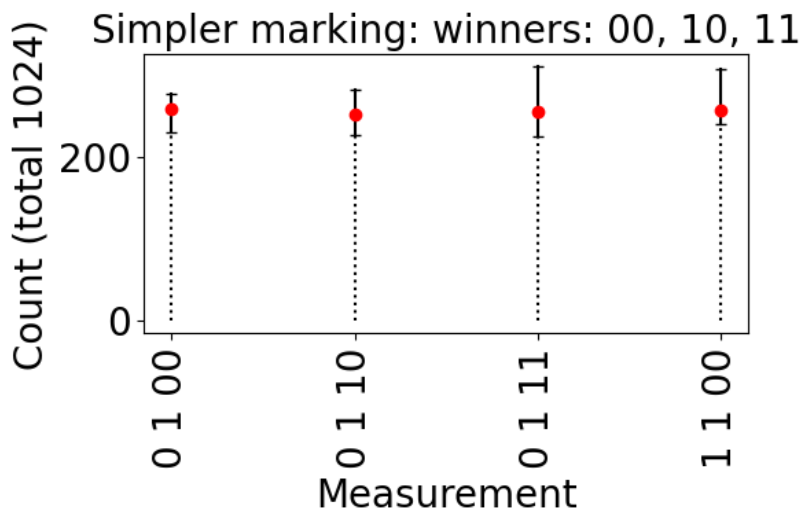}
&
\includegraphics[width=0.24\textwidth]{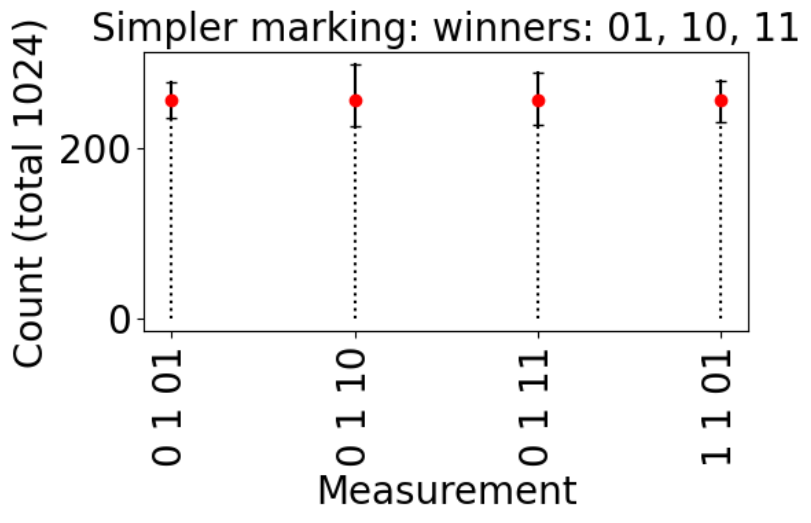}
&
\includegraphics[width=0.24\textwidth]{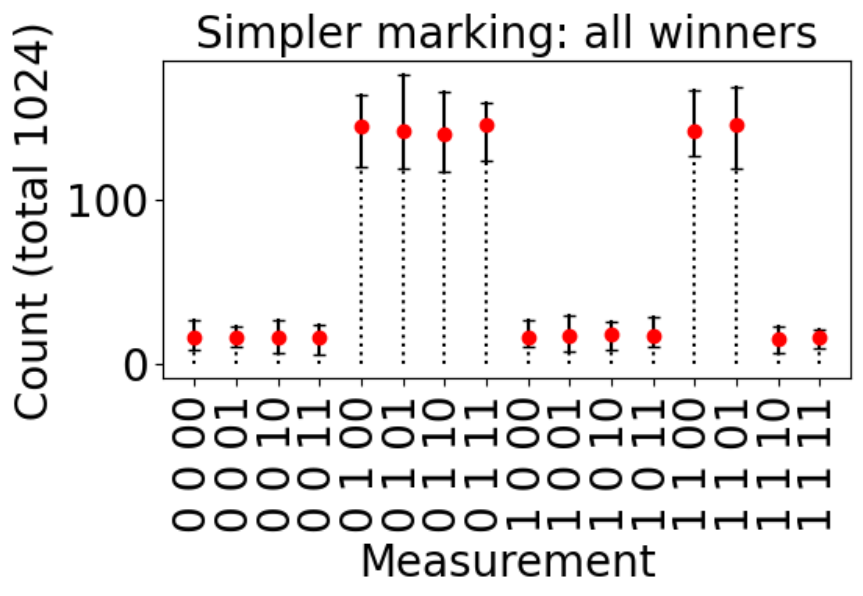}
\\
\end{tabular}
\caption{Plots of experimental results on a two-qubit system simulation. A target state has prefix \texttt{01}.}
\label{fig: two-qubit plots}
\end{figure}

Some arbitrarily selected examples of experimental results are shown in Figure~\ref{fig: two-qubit plots}.
Note that simpler Eigenmarking has its answer(s) 
with tag $ty=$ \verb|01| (prefix \verb|01| in plots).
States with no count are omitted from the plots,
so some plots may appear to show fewer states.
The plot title specifies a scenario: a number of winners representing a number of answers.
The max, median, and min counts are shown: red dot represents median of counts from 40 repeated experiments.

\subsection{Entailment model checking example}
\begin{figure}[htbp]
    \begin{tabular}{ccc}
    \includegraphics[width=0.3\textwidth]{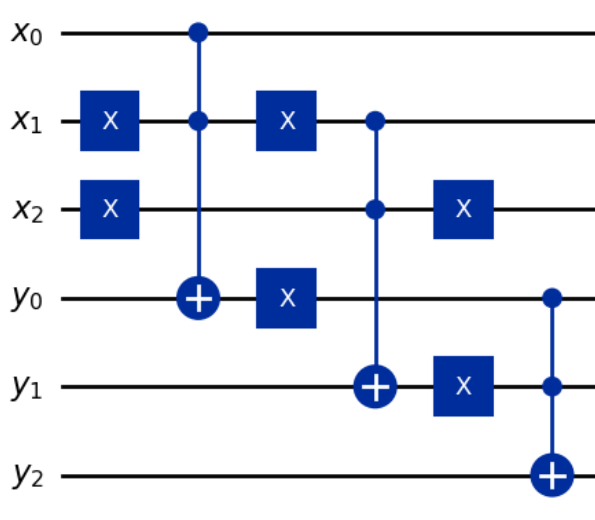}
    &
    \includegraphics[width=0.3\textwidth]{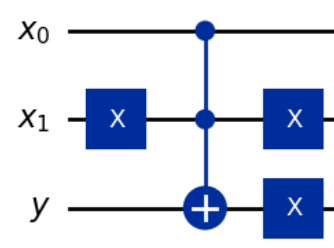}
    &
    \includegraphics[width=0.3\textwidth]{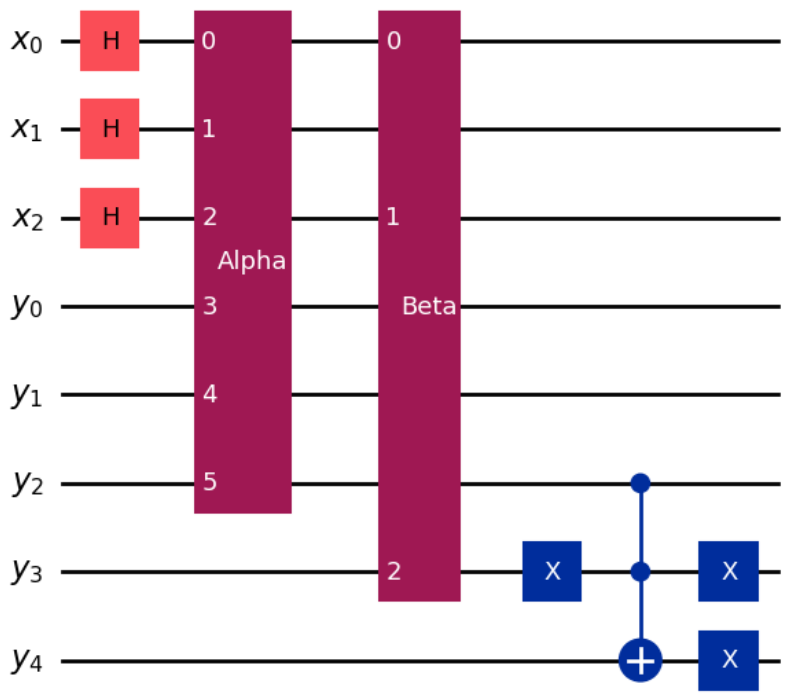}
    \end{tabular}
    \caption{Diagrams: knowledge $\alpha: \{A \Rightarrow B, B \Rightarrow C\}$ (left), sentence $\beta_1: A \Rightarrow C$ (middle), and entailment checking (right).}
    \label{fig: entailment checking}
\end{figure}

\begin{figure}[htbp]
    \begin{tabular}{cccc}
    \includegraphics[width=0.24\textwidth]{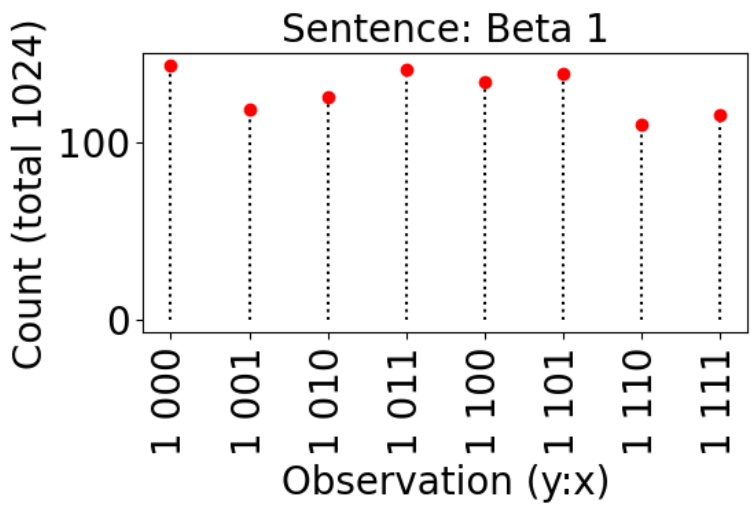}
    &
    \includegraphics[width=0.24\textwidth]{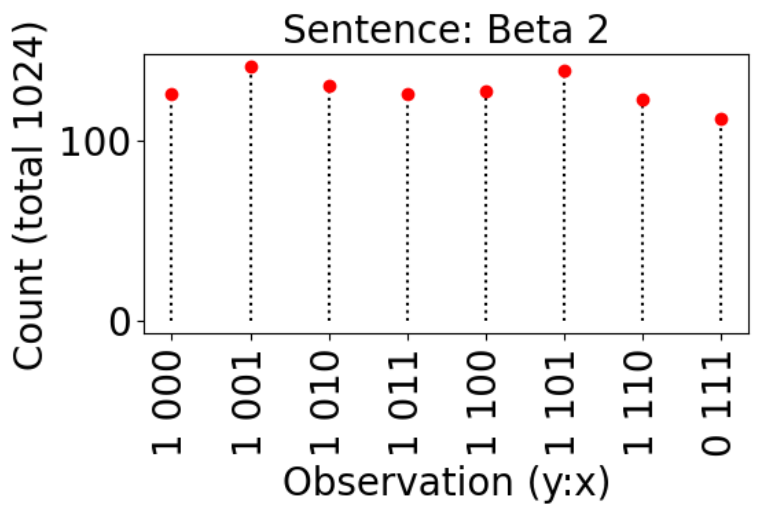}
    &
    \includegraphics[width=0.24\textwidth]{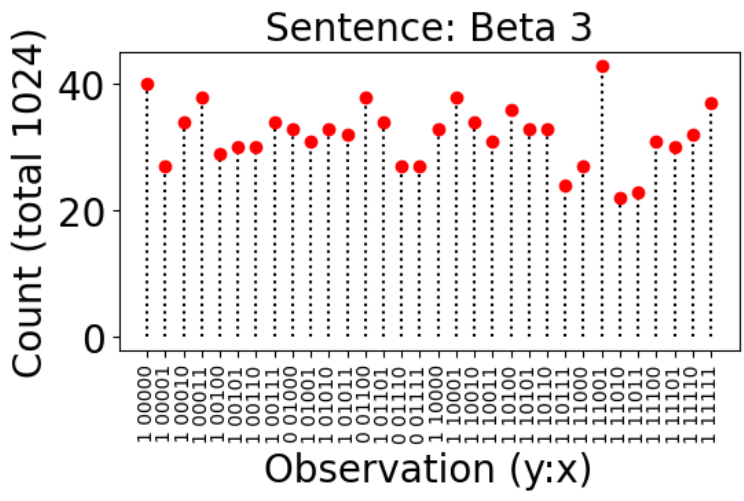}
    &
    \includegraphics[width=0.24\textwidth]{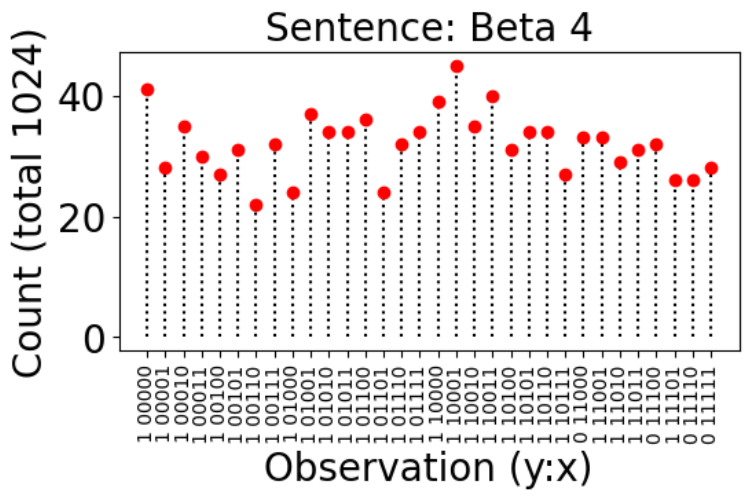}    
    \end{tabular}
    \caption{Entailment checking simulation. The decisive logic is shown as the first digit, 
    e.g., $\beta_1$ has no violation: all $y$'s is $1$'s.}
    \label{fig: entailment checking sim}
\end{figure}

Figure~\ref{fig: entailment checking} shows diagrams implemented entailment model checking for an example described in \textsection~Background.
The simulation results are shown in Figure~\ref{fig: entailment checking sim}. Note that $\beta_3$ and $\beta_4$ introduces more logical symbols, hence more states.

\end{document}